\documentclass[twocolumn,showpacs,preprintnumbers,a4,prb]{revtex4}
\usepackage{graphicx}%
\usepackage{dcolumn}
\usepackage{amsmath}
\usepackage{amssymb}
\usepackage{bm}

\makeatletter
\def\btt#1{\texttt{\@backslashchar#1}}%
\DeclareRobustCommand\bblash{\btt{\@backslashchar}}%
\makeatother

\begin{document}

\title{Orthorhombic crystalline field splitting on orbital and magnetic
       orders in KCuF$_3$}

\author{Dong-Meng Chen$^{1,2}$ and Liang-Jian Zou$^{1}$}
\affiliation{\it  1~ Key Laboratory of Materials Physics, Institute of 
    Solid State Physics, Chinese Academy of Sciences, P. O. Box 1129, 
    Hefei 230031, China}
\affiliation{\it  2~ Graduate School of the Chinese Academy of Sciences}
\date{\today}

\begin{abstract}
   Magnetic and orbital structures in KCuF$_{3}$ are revisited 
by the cluster self-consistent field approach developed recently. 
We clearly showed that due to the inherent frustration, 
the ground state of the system with the superexchange and Jahn-Teller 
phonon-mediated orbital couplings is highly degenerate without 
broken symmetry; the orthorhombic crystalline field splitting 
arising from static Jahn-Teller distortion stabilizes the orbital 
ordering, about 42$\%$ in the $x^{2}-y^{2}$ orbit and 58$\%$ in the 
$3z^{2}-r^{2}$ orbit in sublattices. The magnetic moment of Cu is 
considerably reduced to 0.49$\mu_{B}$, and the magnetic coupling 
strengths are highly anisotropic, J$_{c}$/J$_{ab}$ $\approx$ 26. 
These results 
are in agreement with the experiments, implying that as an orbital 
selector, the crystalline field plays an essential role in stabilizing 
the ground state of KCuF$_{3}$. The 1s-3d resonant X-ray scattering 
amplitudes in KCuF$_{3}$ with the {\it type-a} and {\it type-d} 
structures are also presented.

\end{abstract}
\pacs{71.27.+a, 75.25.+z, 71.70.Ch, 71.70.Ej}
\maketitle

\section{Introduction}

    Three-dimensional pseudocubic perovskite KCuF$_{3}$ has attracted 
extensive interest since 1970s for its unusual low-dimensional 
antiferromagnetic (AFM) and orbital ground state. In this compound the 
Cu$^{2+}$ ion has 3$d^9$ configuration with the fulfilled $t_{2g}$ 
orbits and the twofold-degenerate e$_{g}$ orbits occupied by one hole,
the latter leads to the Jahn-Teller (JT) effect. 
The orbital degree of freedom of the holes interplays with spin and 
lattice degrees of freedom, which results in the orbital regular 
occupation in real space, i.e. the orbital ordering.
The orbital polarization of the hole is usually depicted by a pseudospin 
operator, $\vec{\tau}=1/2 \sum_{ab}c^{\dag}_{a} 
{\bf{\sigma}}_{ab} c_{b}$, here $c^{\dag}_{a}$ creates a hole at orbit 
$a$, and $\sigma$ denotes the {\it Pauli} matrix. 
$\tau_{i}^z=1/2$ represents the full orbital polarization in 
$|3z^2-r^2\rangle$ and $\tau_{i}^z=-1/2$ the orbital 
polarization in $|x^2-y^2\rangle$. The orbital polarization 
degree represented by $\langle\tau_{i}\rangle$ is called 
the {\it orbitalization}.
Up to date, the orbital ordering is considered as the essential factor 
in stabilizing the abnormal magnetic structure in KCuF$_{3}$, and is 
observed recently by the resonant x-ray scattering (RXS) experiment
\cite{Caciuffo02, Paolasini02}. 
Though it has been known very early that the combination of the 
electronic superexchange (SE) interaction and the JT effect is 
responsible for the orbital ordering in KCuF$_{3} $\cite{Kugel73}, 
as shown in the following, the uniqueness of the orbital ground state 
and the magnetic and the orbital experimental results available are 
not well understood theoretically \cite{Kugel73, 
Liechtenstein95,Medvedeva02,Kanamori60,Englman71, Kataoka01}.

    In pseudocubic perovskite KCuF$_{3}$ the adjacent CuF$_{6}$ 
octahedra in ({\bf a,b}) plane are elongated along the {\bf a} and 
{\bf b} axis alternatively due to the JT distortion. 
There are two tetragonal crystal polytypes discriminated as {\it type-a} 
by the antiferro-distortion stacking the {\bf ab}-planes 
and {\it type-d} by the ferro-distortion stacking the {\bf ab}-planes
\cite{Okazaki61,Buttner90}. 
The RXS results \cite{Caciuffo02, Paolasini02} showed that in KCuF$_{3}$ 
the ground state is the G-type antiferro-orbital (AFO) order for 
{\it type-a} structure; and the C-type AFO arrangement for {\it type-d} 
structure. In these two structures the Cu spins are A-type AFM 
order below 39 K for {\it type-a} and below 22K for {\it type-d}; 
in both situations the averaged magnetic moment of each Cu ion is about 
0.49 $\mu_{B}$ at 4 K \cite{Hutchings69}. The magnetic couplings in 
KCuF$_{3}$ are highly anisotropic, and the neutron scattering 
experiment \cite{Satija80, Hutchings69} gave rise to J$_{ab}$/J$_{c} 
\sim$ -0.01, suggesting significant one-dimensional character.

   More than thirty years ago Kugel and Khomskii\cite{Kugel73} proposed 
a SE model from the pure electron-electron interaction, the so-called 
{\it K-K model}, between Cu 3d electrons to describe the role of 
orbital degree of freedom in the A-type AFM structure; by comparing 
the energies of different magnetic structures in the classical 
approximation, they found the A-type 
AFM is stable; while for the orbital configuration, the G-AFO ordering
is degenerate with the C-type AFO ordering. 
Further they showed \cite{Kugel82} that the anharmonic distortion from 
the JT effect lowering the lattice symmetry might give rise to the 
orbital ordering. 
In fact the orbital part of the {\it K-K model} is inherently frustrated 
\cite{Khomskii03}; Feiner {\it et al.} \cite{Feiner97, Oles00}
pointed out that due to quantum spin-orbital wave excitations, 
this frustrated SE interaction leads to a spin-orbital liquid state; 
and when a tetragonal crystalline field (CF) splitting, E$_{z}$, is 
applied to the K-K model, the ground state is ferro-orbital at large 
E$_{z}$.
Some other authors \cite{Englman71} emphasized that the phonon-mediated 
orbital coupling arising from the cooperative JT effect gives 
rise to the orbital ordering. However, 
Khomskii and Mostovoy \cite{Khomskii03} recently pointed out that this 
effective orbital-orbital interaction is also inherently frustrated,
similar to the orbital part in the electronic SE interaction.
In this case the JT orbital interaction combining the SE 
interaction favors the para-orbital or orbital liquid phase
since the quantum fluctuations of the pseudospins $\vec{\tau}$ are still
very large.

  Under the cooperative JT distortion, the low-temperature 
crystal structure of KCuF$_{3}$ is tetragonal; meanwhile, the local 
crystalline field of CuF$_{6}$ arising from the static Jahn-Teller 
distortion is orthorhombic. The effect of the 
orthorhombic CF splitting on the orbital order and the magnetic order 
of KCuF$_{3}$ was seldom taken into account in the past literatures. 
In the recent $ab~ initio$ study within the LDA+U scheme, Binggeli and 
Altarelli \cite{Binggeli04} found that in the {\it type-a} JT distorted
tetragonal structure, the orbital ordering is G-type AFO and the orbital 
RXS intensity agrees with the experimental observation; in contrast,
Medvedeva {\it et al.} \cite{Medvedeva02} found in the absence of the JT 
distortion, the C-type AFO ground state is degenerate with G-type AFO. 
Therefore the realistic orthorhombic CF splitting is crucial for the 
stable orbital-ordered ground state.
Nevertheless, the $ab~ initio$ study underestimates the spin and orbital 
quantum fluctuations in KCuF$_{3}$. It deserves to explore in detail on 
how the spin and orbital quantum fluctuations and the role of 
orthorhombic CF splitting on the orbital ordered ground state.

In this paper, after deriving the orthorhombic CF splitting, we study 
the combination effect of the electronic SE coupling, the effective 
orbital JT coupling and the CF splitting on the 
orbital and spin ground state. Utilizing the cluster self-consistent 
field (Cluster-SCF) approach \cite{Zou05}, we demonstrated that 
the orthorhombic CF splitting plays a key role in stabilizing the 
orbital ordered phases of KCuF$_{3}$ in {\it type a} and {\it type d} 
structures: the orbital and spin fluctuations considerably reduce 
the magnetic moment of Cu spin to 0.49$\mu_{B}$, and the strong 
anisotropy in spin correlation functions and orbital correlation 
functions results in the ratio of magnetic coupling strengths, 
J$_{c}$/J$_{ab}$ $\approx$ 26. 
The azimuthal dependence of the RXS intensity is also calculated for 
the $1s-3d$ excitation event. 
The experimental results could be consistently understood in the 
present theory.   
The rest of this paper is organized as follows: in Sec.II we describe  
the effective Hamiltonian and the Cluster-SCF method; then 
we present the theoretical results and discuss the role of CF in 
magnetic and orbital orderings in Sec.III; 
the azimuthal angle dependence of the RXS intensity is given in Sec.IV;
and the last section is devoted to the remarks and summary.


\section{Model Hamiltonian and Cluster-SCF Method}
   According to the preceding analysis, the effective Hamiltonian of 
KCuF$_{3}$ contains three parts
\begin{equation}
     H=H_{SE}+H_{JT}+H_{CF}
\end{equation}
The first term $H_{SE}$ represents the electronic SE coupling between 
two nearest-neighbor (N.N) $e_{g}$ holes of Cu$^{2+}$ ions derived from 
the generalized twofold degenerate Hubbard model \cite{Castellani78},
\begin{eqnarray}
    H_{SE} &=& \sum_{\substack{\langle ij\rangle_{l}\\l=x,y,z}}
(J_{1}\vec{s}_{i}\cdot\vec{s}_{j_{l}}
       +J_{2}I_{i}^l\vec{s}_{i}\cdot\vec{s}_{j_{l}}
       +J_{3}I_{i}^lI_{j_{l}}^l\vec{s}_{i}\cdot\vec{s}_{j_{l}}
\nonumber\\
       && ~~~~~~~~~+ J_{4}I_{i}^lI_{j_{l}}^l )
\end{eqnarray}
where the constants $J_{1},~ J_{2},~ J_{3}$ and $J_{4}$
are the superexchange coupling strengths:
\( J_{1}=8t^2\left[U/({U}^2-J_{H}^2)-J_{H}/(U_{1}^2-J_{H}^2)\right] \),
\( J_{2}=16t^2\left[1/(U_{1}+J_{H})+1/(U+J_{H})\right] \),
\( J_{3}=32t^2\left[U_{1}/(U_{1}^2-J_{H}^2)-J_{H}/(U^2-J_{H}^2)\right] \), 
and \( J_{4}=8t^2\left[(U_{1}+2J_{H})/(U_{1}^2-J_{H}^2)+
J_{H}/(U^2-J_{H}^2)\right] \), respectively. $U$ and $U_{1}$ 
are the intra- and inter-orbital Coulomb interactions, and $J_{H}$
is the Hund's coupling. Due to the $pd$ hybridization between Cu $3d$ 
and F $2p$ orbitals \cite{Binggeli04}, we take the relationship 
$U=U_{1}+J_{H}$.
The parameters U=7.5 $eV$ and J$_{H}$=0.9 $eV$ are adopted from
the constrained LDA computation for KCuF$_{3}$\cite{Liechtenstein95}. 
The hopping integral along the $c$ direction is the largest, 
t$_{3z^2-r^2,3z^2-r^2}$ = $4t$, we take $t$= 0.12 eV, thus the energy 
scale of the superexchange coupling is $J=16t^{2}/U = 30.7~ meV$.
$\vec{s}_{i}$ denotes the spin at site i, while the operator 
$I_{i}^l=\cos\left(2\pi m_{l}/3\right)\tau_{i}^z-\sin\left(2\pi 
m_{l}/3\right)\tau_{i}^x$, the index $l$ denotes the direction of a 
bond, $l= x, y$ or $z$, corresponding to the crystal axes, $a,b$ and 
$c$; $\langle ij\rangle_{l}$ connects site $i$
and its nearest-neighbor site $j$ along the $l$ direction, and
$(m_{x}$,$m_{y}$,$m_{z})$=$(1$,$2$,$3)$.
The orbital pseudospin operators $\tau_{i}^z$ and $\tau_{i}^x$ are
the components of $\vec{\tau}$.

The second term in Eq.(1) represents the phonon-mediated orbital coupling
from the cooperative JT effect. Through eliminating the phonon operator, 
one can get an effective orbital-orbital interaction H$_{JT}$
\cite{Englman71, Kugel82, Khomskii03},
\begin{equation}
H_{JT}=g_{JT}\sum_{\substack{\langle ij\rangle_{l}\\l=x,y,z}}
I_{i}^lI_{j_{l}}^l
\end{equation}
It has the same form as the orbital part of the last term in Eq.(2), 
both of them contribute to the orbital frustration and quantum 
fluctuations. The Jahn-Teller distortion energy E$_{JT} \sim 130$ meV 
\cite{Caciuffo02}, g$_{JT}$ is the same order as E$_{JT}$ in magnitude.

   The static JT effect distorts the symmetry of CuF$_{6}$ octahedra to 
orthorhombic, in which there are three different Cu-F bonds, the medium 
length Cu-F bond is along the z-axis, the long and short Cu-F
bonds alternate along the $x$ and the $y$ axes. All of the F-Cu-F angles 
are 90$^0$ or 180$^0$, and no rotation is found \cite{Buttner90}. 
Employing the point charge model to calculate the CF splitting, we obtain
the orthorhombic CF splitting of the holes,
\begin{equation}
H_{CF}=\sum_{i}\left(V_{iz}\tau^{z}_{i}+{V_{ix}\tau^{x}_{i}}\right)
\end{equation}
The intermediate compression of the Cu-F bond along c-axis implies the 
sign of $V_{iz}$ is negative for the hole, $V_{iz}<0$, since the 
compressed octahedra lifts the $|3z^2-r^2\rangle$ orbit, thus the 
$|3z^2-r^2\rangle$ orbit is in favor of hole occupation; and the 
component $V_{ix}$ mixes the 
two $e_{g}$ orbitals at site $i$. The former favors of the 
orbital ordering, while the latter tends to destroy the orbital ordering.
Using the crystal structure data in Ref.10 we estimate the ratio 
$|V_{x}/V_{z}|\sim 2$, and $V_{z}$ 
is the same order in magnitude as the superexchange interaction.

  It is a huge challenge to treat the spin-orbital correlations and 
fluctuations and to find the ground state of such strongly correlated 
systems with high accuracy. To study the groundstate properties, we 
apply the Cluster-SCF approach\cite{Zou05} developed recently to 
deal with the complicated spin-orbital Hamiltonian (1). This 
approach combines the exact diagonalization for a central cluster 
and the self-consistent field for surrounding atoms. 
The main idea of the approach is described as follows: consider
a proper cluster, usually the unit cell of the compounds, in which the 
3d electrons interact via Eq.(1). First, we substitute the spin coupling 
$\vec{s}_{i}\cdot\vec{s}_{j}$ in the cluster Hamiltonian, H, with the 
spin correlation functions $\langle\vec{s}_{i}\cdot\vec{s}_{j}\rangle$, 
here
\begin{eqnarray}
    H  &=& \sum_{\substack{\langle ij\rangle_{l}\\l=x,y,z}} \left[ 
F\left(\vec{\tau}_{i},\vec{\tau}_{j_{l}}\right)
\vec{s}_{i}\cdot\vec{s}_{j_{l}}+(J_{4} +g_{JT})I_{i}^lI_{j_{l}}^l\right] 
\nonumber \\
&& ~~~~~~~~  + 
\sum_{i}\left(V_{iz}\tau^{z}_{i}+{V_{ix}\tau^{x}_{i}}\right),
\end{eqnarray}
with
\begin{eqnarray}
   F\left(\vec{\tau}_{i},\vec{\tau}_{j}\right) = J_{1} +
    \frac{J_{2}}{2}(I_{i}^l+I_{j_{l}}^l) +J_{3}I_{i}^lI_{j_{l}}^l .
\nonumber 
\end{eqnarray}
And then diagonalize the orbital part of the cluster Hamiltonian in the 
presence of the orbital SCF \cite{Chen05}, hence obtain the 
orbitalization $\langle\vec{\tau}\rangle$ and the orbital correlation 
functions $\langle\vec{\tau}_{i}\cdot\vec{\tau}_{j}\rangle$.
Second, substitute the orbital operator, $\vec{\tau}$, and their 
coupling $\vec{\tau}_{i}\cdot\vec{\tau}_{j}$ with the 
orbitalization and the orbital 
correlation functions obtained, and diagonalize the spin part of the 
cluster Hamiltonian in the presence of the spin SCF, thus obtain a new 
set of spin correlation functions. Repeat the above steps iteratively
until the groundstate energy, the spin and the orbital 
correlation functions self-consistently converge to the accuracies. 
The advantage of the present approach superior 
to the traditional mean-field method is that the short-range spin and 
orbital correlations and quantum fluctuations are taken into account.
In comparison with the traditional mean-field method, one can obtain 
more better results for some simple models with large quantum 
fluctuations, such as the Heisenberg AFM model \cite{Chen05}.

\section{Results and discussions}

    In this section we first investigate the ground state of KCuF$_{3}$
under the electronic SE interaction and the JT phonon-mediated orbital 
interaction, then explore the role of orthorhombic CF splitting in the 
ground state.

\subsection{Superexchange and JT Orbital Interactions}

  For the electronic SE coupling in Eq.(2), the mean-field results 
\cite{Kugel73,Medvedeva02} 
suggested the staggered orbital order with x$^{2}$-z$^{2}$/
y$^{2}$-z$^{2}$ orbits. However, utilizing the cluster-SCF approach,
we find the spin-orbital ground state is composed of numerous 
degenerate states: such as the {\it Neel} AFM order with ferro-orbital  
structure, in which the hole occupies one of the three orbits 
among the $|3x^2-r^2\rangle$, the $|3y^2-r^2\rangle$ and the 
$|3z^2-r^2\rangle$ orbits; or the {\it Neel} AFM order with the
alternating plaquette valence-bond order, in which one plane, 
for example, the xy-plane, is 
occupied by the $|3x^2-r^2\rangle$ orbit and the nearest 
neighbor ones is taken up with the $|3y^2-r^2\rangle$ orbit, etc.
The degenerate ground state contains the alternating plaquette 
valence-bond component, in agreement with Feiner {\it et al.}'s result 
\cite{Feiner97} by considering Gaussian quantum fluctuations. 
This also confirms the validity and the efficiency of our Cluster-SCF 
approach.  
In this case, the A-type AFM structure observed in realistic KCuF$_{3}$ 
is not the candidate of the ground state. 
On the other hand, the nearest-neighbor orbital correlation functions
are significantly different from zero. The short-range orbital 
correlations are strong, indicating that the system with only the 
electronic SE interaction is a spin-orbital liquid phase.

   In such a spin-orbital correlated system, the spin alignment strongly 
depends on the orbital configuration. In fact one could easily find that 
the system with the SE interaction is a  
spin-orbital liquid phase, rather than a spin-orbital ordered state. 
Due to the spin rotation symmetry, the spin interaction in the SE 
coupling is Heisenberg-like. After averaging over orbital freedom degree,
the spin exchange coupling strength along the $l$-axis in Eq.(5) reads
\begin{eqnarray}
    J^{l}_{s}&=&\langle F(\vec{\tau}_{i}^l,\vec{\tau}_{j}^l)\rangle 
\nonumber \\
             &=&J_{1}+\frac{J_{2}}{2}\left(\langle I_{i}^l\rangle+
                \langle I_{j}^l\rangle\right)+
              J_{3}\langle I_{i}^lI_{j}^l\rangle,  \nonumber
\end{eqnarray}
which is isotropic for the x-, the y- and the z-axes, such an 
interaction does not lead to the anisotropic A-type AFM 
structure in KCuF$_{3}$, unless the orbital symmetry is broken.
In the orbital part of the SE interaction in Eq.(2), after averaging 
over the spin coupling, the orbital part along the $l$-direction 
reads,
\begin{equation} 
  h^{l}_{o} = \frac{J_{2}}{2}\langle\vec{s}_{i}\cdot\vec{s}_{j_{l}}
      \rangle \left(I_{i}^l+I_{j_{l}}^l\right)+
        \left(J_{3}\langle\vec{s}_{i}\cdot\vec{s}_{j_{l}}\rangle
        +J_{4}\right)I_{i}^lI_{j_{l}}^l,   
\end{equation} 
The first term is similar to a magnetic field, called the orbital field.
This orbital field is frustrated for the orbital occupation,
which favors either $|3z^2-r^2\rangle$ or $|x^2-y^2\rangle$ orbits 
along the $z$-axis; while along the $x$ direction, it favors
either $|3x^2-r^2\rangle$ or $|y^2-z^2\rangle$ orbits.
Unfortunately the orbital interaction part, $I_{i}^lI_{j_{l}}^l$, in 
Eq.(6) is also inherently frustrated: no matter what the 
coefficient of the second term in Eq.(6) is, negative or positive, 
this orbital correlation favors a frustrated ground state, as 
pointed by Khomskii et al. \cite{Khomskii03} 
On the one hand, if the coefficient of $I_{i}^lI_{j_{l}}^l$ is 
negative, the orbital exchange coupling favors different ferro-orbital 
occupations along different directions, in accordance with the 
frustration of the orbital field mentioned above; on the other 
hand, if the coefficient is positive, the orbital polarizations in 
each bond arising from the orbital field are opposite to that 
arising from the orbital coupling, which enhances the frustration 
effect.

   One notices that if the spin correlation functions are strongly 
anisotropic, the frustration effect is greatly suppressed. For example, 
as the spin correlation along the $z$-axis is so strong that
$\langle \vec{s}_{i}\cdot\vec{s}_{j_{z}}\rangle\approx -3/4$,
while the spin correlation functions along the $x,y$-axes almost vanish,
 $\langle \vec{s}_{i}\cdot\vec{s}_{j_{x,y}}\rangle\approx 0$,
then the orbital part of the SE interaction becomes  
\begin{eqnarray}
  h_{o} = J_{4} \sum_{i}[ I_{i}^xI_{i+x}^x+I_{i}^yI_{i+y}^y-
0.95I_{i}^zI_{i+z}^z -1.78I_{i}^z ]
\end{eqnarray}
the $x,y$-components of the orbital field approach to zero, leaving  
a large $z$-component in Eq.(7). Obviously the strong uniaxial orbital
field suppresses the frustration of the 
orbital exchange coupling, and singles out the $|3z^2-r^2\rangle$ 
orbit in each site, forming the ferro-orbital order; our numerical 
calculation confirms this result. 
Also it could be shown that strongly anisotropic plaquette-valence-bond 
correlation of the spins favors the alternating plaquette-valence-bond 
orbital order. As we will shown later, the highly anisotropic magnetic 
correlation is the consequence of the orbital ordering in KCuF$_{3}$.

   Neglecting the spin and orbital quantum fluctuations, Kugel 
and Khomskii
\cite{Kugel73} found the mean-field solution of the SE interaction is 
the A-type AFM and G-type/C-type AFO order. In fact, their classical 
approximation to the spins as the A-type AFM order introduced  
an orbital field (-$\sum_{i}I_{i}^z$). This orbital field 
lifts most degenerate states of the orbital ground state, leaving 
the G-type and the C-type AFO configurations as the candidates;
the fourfold rotation symmetry of the spin structure also confines the 
possible orbital ground state as the G-type or C-type AFO structure, in 
agreement with Goodenough-Kanamori empirical rule \cite{Goodenough63}. 
Nevertheless, the orbital field do not remove the degeneracy of the 
G-type and the C-type AFO structures, as shown in Ref.3.
However, as found in the experiments \cite{Hutchings69}, the 
average magnetic moment of each Cu spin is about 0.49 $\mu_{B}$, 
only a half of the classical expectation, suggesting that spin 
quantum fluctuations in KCuF$_{3}$ is very strong, and the 
classical approximation to the spins is not appropriate.

  The phonon-mediated orbital-orbital coupling with discrete cubic 
symmetry, H$_{JT}$ in Eq.(3), is also strongly frustrated. It has 
the same impact on the orbital ground state as the second term in 
Eq.(6) does. Early treatment \cite{Englman71} to H$_{JT}$ by the 
mean-field approach is questionable since the presumed orbital order was 
equivalent to introduce an artificial orbital field to break the orbital 
symmetry, however the orbital field did not really exist.
We notice that H$_{JT}$ can be absorbed in the second term in Eq.(6),
so the combination of the JT orbital coupling H$_{JT}$ and the SE 
interaction, H$_{SE}$, does not break the symmetry of spin and 
orbital in cubic crystal structure. Therefore in the absence of 
the crystalline field splitting from the static JT distortion, the 
ground state of the system is an orbital liquid or para-orbital phase.    
Accordingly, the orbital symmetry in KCuF$_{3}$ should be broken by 
the static JT distortion, i.e. the orthorhombic CF splitting, H$_{CF}$.
In what follows we explore the role of orthorhombic CF splitting in 
the ground state in KCuF$_{3}$.

\subsection{Role of Orthorhombic Crystalline Field}

   In perovskite KCuF$_{3}$ there exist two kind cooperative JT 
distortions at low temperature\cite{Okazaki61,Buttner90}. With 
respect to these two different small lattice 
distortions, KCuF$_{3}$ exhibits two slightly different crystalline 
phases, the {\it type-d} and the {\it type-a} structures.
The orbital ground state in the {\it type-a} structure differs 
from that in the {\it type-d} structure due to different orthorhombic 
CF splittings of the CuF$_{6}$ octahedra.

   We first study the orbital components of single Cu atom under the 
orthorhombic CF. For $|V_{x}/V_{z}| = 2$, the orbital 
wavefunctions always consist of two orbital patterns, a low energy 
pattern $|a\rangle=0.851|3z^2-r^2\rangle\pm 0.526|3x^2-y^2\rangle$,
in which the hole completely occupies this pattern at large $|V_{z}|$,
and a high energy pattern $|b\rangle=0.526|x^2-y^2\rangle\mp 0.851|3z^2-
r^2\rangle$; here $'\pm'$ refer to the two sublattices of the 
antiferro-distortion in the two crystalline phases. Approximately, 
$|a\rangle\approx|y^2-z^2\rangle$ and $|b\rangle\approx 
|3x^2-r^2\rangle$ for '+'; and $|a\rangle\approx|x^2-z^2\rangle$ 
and $|b\rangle\approx |3y^2-r^2\rangle$ for '-'.
Such combinations are also consistent with the orthorhombic distortions 
of the CuF$_{6}$ octahedra: 
if the Cu-F bond is elongated along the x axis, the CF singles out the 
$|y^2-z^2\rangle$ orbit, corresponding to '+'; on the other hand, if 
Cu-F bond is elongated along y direction, the energy of 
$|x^2-z^2\rangle$ orbit is lower, corresponding to '-'.

   In the lattice case, the orthorhombic CF splitting competes with the 
SE interaction and the JT orbital coupling, the orbital occupation 
of Cu 3d holes depends not only on the CF splitting ratio 
$|V_{x}/V_{z}|$, but also on the magnitude of $V_{z}$. When 
$|V_{z}|$ is very small, due to the orbital frustration and large 
quantum fluctuations, the orbital symmetry is not broken, and the 
ground state of the system is still an orbital liquid or 
para-orbital phase. The critical value of $|V_{z}|$ breaking the 
orbital symmetry relies on the JT orbital coupling.  
For large $|V_{z}|$, the transverse CF splitting V$_{x}$ alternating 
in the $xy$-plane in the {\it type-d }structure gives rise to C-type AFO 
configuration; 
as a contrast in the {\it type-a} structure, the staggered transverse CF 
splitting V$_{x}$ in the $x,~y~ and~z $ directions gives rise to 
G-type AFO configuration. At $|V_{z}|$=0.5J, 
the sublattice orbitalization and the orbital correlation functions  
listed in Table I definitely show the G-type AFO correlation in 
the {\it type-a} structure and the C-type AFO correlation in the 
{\it type-d} structure. Some magnetic properties in both structures 
are also collected in Table I.
Therefore the presence of the orthorhombic CF breaks the 
discrete orbital symmetry, suppresses the orbital frustration and quantum
fluctuation, and establishes the long-range orbital order in KCuF$_{3}$.

   As soon as the orthorhombic CF singles out the orbital 
structure, it also stabilizes the magnetic structure simultaneously, 
hence the spin-orbital ground state. Under the full Hamiltonian, 
our numerical results show that 
the magnetic ground states are the A-type AFM order both for the 
{\it type-a} and for the {\it type-d} crystalline phases.
And the spin correlations are strongly anisotropic, 
$\langle \vec{s}_{i}\cdot\vec{s}_{j_{z}}\rangle/\langle
\vec{s}_{i}\cdot\vec{s}_{j_{x,y}}\rangle \approx 10$. Furthermore, 
we obtain the spin coupling strengths J$_{z}$ and J$_{x,y}$, which 
are 16.6 meV and 0.64 meV, respectively, giving rise to
$|J_{z}/J_{x,y}|$ about 26 for V$_{z}$=-0.5J, as shown in Fig.1.
\begin{figure}[htbp]\centering
\includegraphics[angle=270,width=0.7 \columnwidth]{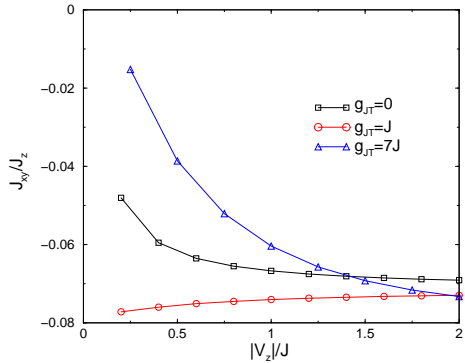}
\caption{Dependence of the ratio of N.N spin coupling strengths 
     J$_{x,y}$/J$_{z}$
     on orthorhombic CF splitting $|V_{z}|/J$. Here $J$ is the energy 
     scale of the SE coupling.}
\label{fig1}
\end{figure}
These results are in agreement with the experimental data in KCuF$_{3}$ 
\cite{Hutchings69,Caciuffo02, Paolasini02}. Such strong anisotropy 
in spin correlations and magnetic couplings is attributed to the 
anisotropic regular distribution of the orbital wavefunctions in 
real space, i.e., the orbital ordering, as we will show later.

  Next we focus on the magnetic moment of each Cu spin. We find that the 
effective magnetic moment of each Cu spin is considerable reduced from 1 
$\mu_{B}$ to 0.496 $\mu_{B}$, the averaged spin of each Cu ion decreases 
to 0.248, about a half of 1/2, consisting with the neutron scattering 
experimental data very well \cite{Hutchings69}. 
The theoretical and experimental results are listed in Table I.
Obviously such great reduction of the magnetic moment arises from the 
spin-orbital quantum fluctuations: one possibility is from the 
low-dimensional AFM spin-wave excitation; the another is from the
spin-and-orbital wave excitation in the spin-orbital system 
\cite{Oles00}, 
which causes more spin flipping via the spin-orbital interaction. 
%
\begin{table}[tbp]
\begin{center}
\caption{Calculated and experimental averaged spin, magnetic couplings,  
         nearest-neighbor spin correlations, orbitalization and 
         orbital correlations. 'a' and 'd' refer {\it type-a} and 
         {\it type-d} phases. J$_{x,y,z}$ are in units of $meV$.
         The theoretical parameters are $g_{JT}=7J$, and $|V_{z}|=0.5J$. }
\vspace{0.6cm}
\begin{tabular}{lccccccc} \hline
    \hline 
    &$\langle s\rangle$  &$\langle J_{x,y}\rangle$  &$\langle 
   J_{z}\rangle$ &$\langle\vec{s}_{i}\cdot\vec{s}_{j}\rangle_{x,y}$ 
  &$\langle\vec{s}_{i}\cdot\vec{s}_{j}\rangle_{z}$  \\ \hline 
  Calc. &0.248  &-0.64 &16.6 &0.070 &-0.684 \\
  Expt. &0.245\cite{Hutchings69}   &-0.2 \cite{Satija80}  &17.5 
   \cite{Satija80}
&      &     \\ \hline
 &$\langle\tau_{z}\rangle$  &$\langle\tau_{x}\rangle$ &$
    \langle\vec\tau_{i}\cdot\vec\tau_{j}\rangle_{x,y}$ 
 &$\langle\vec\tau_{i}\cdot\vec\tau_{j}\rangle_{z}$ \\ \hline
 Calc.$^d$  &0.077 &0.492  &-0.251  &0.249 \\ \hline 
 Calc.$^a$  &0.077 &0.492  &-0.251  &-0.280 \\ \hline
\hline 
\end{tabular}
\end{center}
\end{table}
%
From Table I, one finds the discrepancy between the present ratio of 
$|J_{z}/J_{x,y}|$ ($\approx$ 26) and Satija's experimental fitting data 
($|J_{z}/J_{x,y}|$ $\approx$ 100) \cite{Satija80}, we attribute this
discrepancy to the frozen of the orbital excitations in his fitting to the 
experimental data.
According to the spin exchange couplings shown in Table I, we find that 
in the mean-field approximation, the theoretical {\it Neel} temperature 
of KCuF$_{3}$ is about, 37 K, 
in agreement with the experimental data 39 K in {\it type-a} structure,
confirming that our choice to the theoretical parameters t, U, J$_{H}$, 
g$_{JT}$ and $V_{z}$ is appropriate.

   In fact the influence of the orthorhombic CF splitting and the JT 
orbital coupling on the magnetic moment is not monotonously. As shown 
in Fig.2, the phonon-mediated JT orbital coupling and the 
orthorhombic CF distortion play distinct roles in the magnetic 
moments through affecting the orbital quantum fluctuations and the 
orbital ordering.
At small splitting $|V_{z}|$ and in the absence of the JT coupling, 
$g_{JT}=0$, the orbital field $-I_{i}^z$ in the A-type AFM structure 
is much stronger than that from the CF splitting, resulting in 
large orbital polarization with dominant $|3z^2-r^2\rangle$ orbit; 
with the increase of the CF splitting, the 
transverse CF term V$_{x}$ mixes the $|3z^2-r^2\rangle$ orbit with the 
$|x^2-y^2\rangle$ orbit, leading to the descent of the sublattice 
orbitalization, as seen in Fig.2a.
Meanwhile the decrease of the orbitalization weakens the 
low-dimensionality of the spin correlations, the magnetic moment 
gradually lifts with increasing CF splitting, which can be seen in 
Fig.2b. At sufficient large $|V_{z}|$, 
the orbital occupation is full polarized at the $|a\rangle$ orbit, 
and the magnetic moment saturates to 0.61 $\mu_{B}$ per site.
%
\begin{figure}[htbp]\centering
\includegraphics[angle=270,width=0.7 \columnwidth]{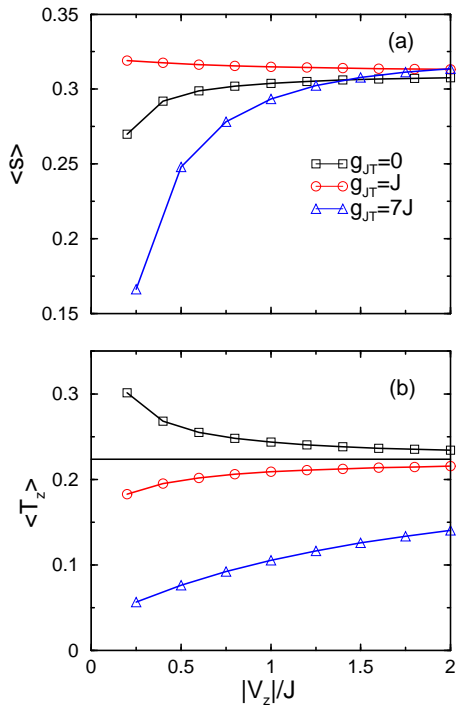}
\caption{Dependence of averaged spin (a) and sublattice orbitalization 
  (b) on orthorhombic CF splitting in different JT orbital coupling.}
\label{fig2}
\end{figure}
%
In this situation, a small CF splitting favoring of the A-type AFM 
structure produces strong local orbital field and orbital 
polarization, suppresses the orbital frustration from the 
superexchange coupling. With the further increase of the CF splitting, 
the transverse component of the CF splitting V$_{x}$ mixes the orbitals
$|3z^{2}-r^{2}\rangle$ and $|x^{2}-y^{2}\rangle$, the orbital  
polarization declines and the anisotropy of orbital and spin 
correlations become weak, 
hence the ratio of the magnetic couplings $J_{z}/J_{x,y}$ decrease with 
the increase for very large CF splitting, as seen in Fig.1.

  When the JT orbital coupling is taken into account, the dependences 
of the spin coupling, the sublattice orbitalization and the magnetic 
moment on $|V_{z}|$ are different for g$_{JT}$/J=1 and for g$_{JT}$/J
=7. The different behaviors arise from the distinct effects of the CF 
splitting on the local orbital field and the frustration term. 
At g$_{JT}$/J=1,  
$|J_{x}/J_{z}|$ is larger than that at g$_{JT}$=0, implying the 
low-dimensional characters of the spin correlations and the spin 
fluctuations become weak. This leads to large magnetic moment 
and small anisotropy, as we find in Fig.1 and Fig.2b. The weakness of 
the anisotropy of the magnetic couplings at g$_{JT}$/J=1 arises 
from the frustration enhancement comparison with that at g$_{JT}$=0.
Our numerical results show that at g$_{JT}$/J=1, the further increase 
in the CF splitting is almost balanced by the orbital field and the 
frustrated orbital-orbital couplings, hence the spin and the orbital 
correlations, the sublattice orbitalization and sublattice magnetization 
and the magnetic coupling almost do not change with the increase of the 
CF splitting, as shown in Fig.1 and Fig.2.

In contrast, the large JT orbital coupling at $g_{JT}=7J$ leads to strong 
orbital fluctuation, hence to small sublattice orbitalization and 
magnetization in Fig.2. In this situation, one would expect more weak 
anisotropy or large ratio $|J_{x}/J_{z}|$; however as shown in 
Fig.1, one curiously finds that the anisotropy is the most strong, 
in comparison with $g_{JT}=0$ and $g_{JT}=J$. We find this strong 
anisotropy may come accidentally in small orthorhombic CF splitting:
in the spin coupling along the $x$-axis, the sign of the constant part 
J$_{1}$ is contrary to that from the other terms, which leads to 
J$_{x}$ be very small, hence small ratio $|J_{x}/J_{z}|$ and large 
anisotropy.
Strong orbital fluctuations also excite spin flipping excitation via the 
spin-orbital coupling. Thus the magnetic moment critically decreases to 
0.49 $\mu_{B}$ at $|V_{z}|/J=0.5$. Further increase of the orthorhombic 
CF splitting greatly suppresses the orbital and spin fluctuations, and 
the sublattice magnetic moment and the orbitalization rise a lot. 
With the further increase of the orthorhombic CF splitting, the 
anisotropy becomes more weak, and the sublattice magnetization 
$\langle S\rangle$ and orbitalization $\langle T_{z}\rangle$ 
further increase gradually.

\section{Resonant X-Ray Scattering for Orbital Order}

   The orbital ordering in KCuF$_{3}$ can be manifested in the RXS peaks 
utilizing the sensitivity of x-ray scattering to the anisotropic density 
of orbital ordered electrons \cite{Caciuffo02,Paolasini02}.
The anisotropy of spatial electronic clouds gives rise to the anomalous 
tensor component in atomic scattering factor, and the 
interference of these atomic scattering amplitudes in the presence of 
long-range orbital order leads to the orbital superlattice reflection
at the structural forbidden position. Obviously, it is more directly
to identify the orbital ordering using the spectral line shapes of the 
quadrupole $1s-3d$ scattering, contrary to the complicated spectra of 
the $1s-4p$ dipole scattering which is often used in present 
experiments \cite{Caciuffo02,Paolasini02}, although 
the signal enhancement of the quadrupole scattering is weaker 
than the dipole scattering\cite{Murakami98,Fabrizio98,Lovesey01}. 
In the following we present the azimuthal angle dependence of the 
$1s-3d$ RXS intensity to directly demonstrate the character of the 
orbital order in KCuF$_{3}$.

  For the C-type AFO ordered ground state with {\it type-d} structure, 
the orbital ordering peaks reflect at
$\left(h,k,l\right)=\left(odd, odd, even\right)$, which are forbidden 
for the structural and magnetic reflections. The sublattice orbital 
wavefunctions consist of two different components: $|\psi_{1}\rangle=
\alpha_{1}|3z^2-r^2\rangle+\alpha_{2}|x^2-y^2\rangle$ and $|
\psi_{2}\rangle=\alpha_{1}|3z^2-r^2\rangle-\alpha_{2}|x^2-y^2\rangle$,
here the coefficients $\alpha_{1,2}$ are the functions of the 
interaction parameters. Then the orbital structural 
factor is read as:
\begin{eqnarray}
  F_{hkl}&=&f\left(\Gamma,r_{2,ds},c,\omega\right)\sqrt{n_{\epsilon k}
     n_{\epsilon'k'}}\alpha_{1}\alpha_{2} \{\epsilon_{z}k_{z}
\nonumber\\
  && \left(\epsilon_{x}^{'}k_{x}^{'}-
\epsilon_{y}^{'}k_{y}^{'}\right)+\left(\epsilon_{x}k_{x}-
    \epsilon_{y}k_{y}\right)\epsilon_{z}^{'}k_{z}^{'} \}
\end{eqnarray} 
where the function $f\left(\Gamma,r_{2,ds},c,\omega\right)$ is the
coefficient depending on the lifetime of the intermediate states,
$\Gamma$, the radial matrix element $r_{2,ds}$, the velocity of
photon c and the incoming photon frequency $\omega$;
$n_{\epsilon\left(\epsilon'\right), k\left(k'\right)}$ is the
density of the incoming (outgoing) beam of photons with
polarization $\vec{\epsilon} $($\vec{\epsilon'}$) and
wavevector $\vec{k} $($\vec{k'}$). 
The azimuthal angle dependence of the RXS intensity is shown in 
Fig.3 at $\left(1,1,0\right)$ reflection for unrotated $\left(
\sigma\sigma'\right)$ and rotated $\left(\sigma\pi'\right)$ channels 
for the perfect $\sigma$ polarized incoming beam. 
\begin{figure}[htbp]\centering
\includegraphics[angle=270,width=0.9 \columnwidth]{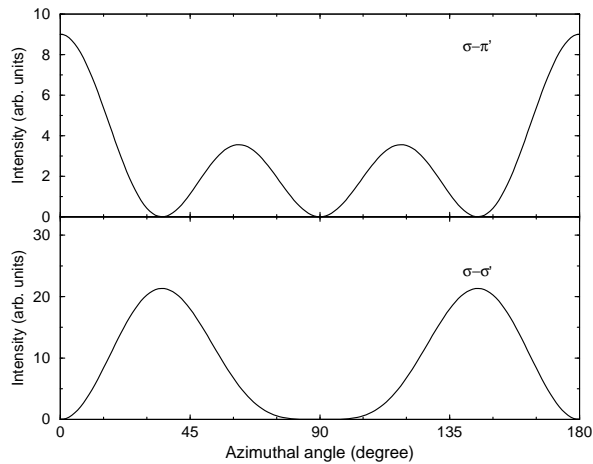}
\caption{The azimuthal angle dependence of the orbital (1,1,0) reflection
         intensity for unrotated $\left(\sigma\sigma'\right)$ and rotated
         $\left(\sigma\pi'\right)$ channels in C-type AFO.}
\label{fig3}
\end{figure}
As a comparison, we also present the azimuthal angle dependence of 
the RXS intensity in Fig.4 for the G-type AFO order 
with {\it type-a} structure at $\left(3,3,1\right)$ reflection. 
\begin{figure}[htbp]\centering
\includegraphics[angle=270,width=0.9 \columnwidth]{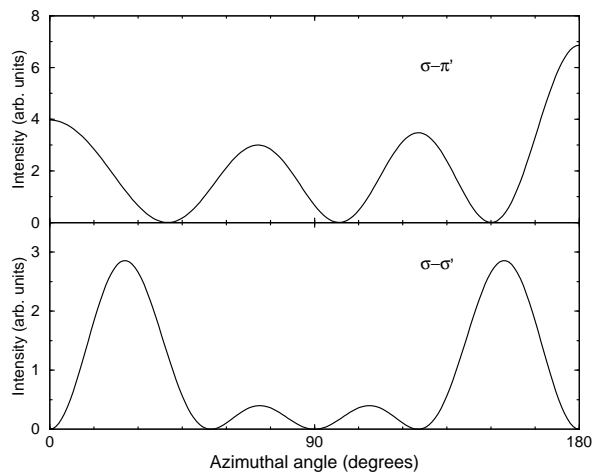}
\caption{Azimuthal angle dependence of the orbital (3,3,1) reflection 
         intensity for unrotated $\left(\sigma\sigma'\right)$ and 
         rotated $\left(\sigma\pi'\right)$ channels in G-type AFO. }
\label{fig4}
\end{figure}
%

  Due to the difference of the orbital superlattice for the C-type 
and the G-type AFO orders, the RXS peaks appear at different orbital 
reflections with $(odd, odd, even)$ for the C-type orbital order in the
{\it type-d} structure, and with $(odd, odd, odd)$ for the G-type 
orbital order in the {\it type-a} structure. These sublattice 
reflections distinguish these orbital orders in different 
crystalline phases of KCuF$_{3}$. 
   The azimuthal angle dependence of the RXS intensities exhibits 
different periods in the {\it type-d} and the {\it type-a} structures, 
as seen in Fig.3 and Fig.4. The periods are $\pi$ and 2$\pi$ for 
the $\sigma-\pi'$ channel in these two structures, respectively; and 
are $\pi/2$ and $\pi$ for the $\sigma-\sigma'$ channel respectively. 
Distinct line shapes of the RXS scattering intensities in the 
$\sigma-\sigma'$ channel also easily identify the two orbital AFO orders.
The different periods are attributed to the different orbital structures 
along the $z$-axis.
One may expect that the 4p orbits of Cu hybridizing with 3d orbits will 
modify the present 1s-3d scattering line shape, but not the period.

\section{Remarks and Summary}

  We notice that in perovskite KCuF$_{3}$, the nonresonant magnetic 
scattering experiment\cite{Caciuffo02} showed that the orbital angular 
momentum {\it L} contribute finite value the total magnetic moment, 
and $L/S \approx 0.29$. 
Obviously the finite orbital moment does not come from the two 
$e_{g}$ orbits, since the expectation of the orbital angular momentum 
$L$ in the orbital basis wavefunctions $|3z^2-r^2\rangle/|x^2-y^2
\rangle$ and any of their combinations is zero.
One possibility of such considerable residual orbital moment is 
attributed to the reduced symmetry of the $t_{2g}$ orbits in 
KCuF$_{3}$ \cite{Hidaka98} or a small fraction of the $t_{2g}$ 
orbit mixing with the $e_{g}$ orbit; another possibility is from 
the hybridization of the $4p$ orbits with the e$_{g}$ orbits. 
In both situations the weak $LS$ coupling in KCuF$_{3}$ will not 
change the spin alignment considerably.

   In the numerous literatures on KCuF$_{3}$, some authors 
\cite{Feiner97, Oles00, Medvedeva02} took the tetragonal CF splitting 
into account to stabilize the A-type AFM order, 
but the degeneracy of the G-type and the C-type AFO orders is not lifted, 
so the ground state under the tetragonal CF is still indefinite. 
Only in the present theory the full consideration of the orthorhombic 
CF splitting, V$_{z}$ and V$_{x}$, together with the JT orbital 
coupling and the SE coupling, can we 
determine the ground state exclusively, and consistently interpret 
the experimental data. Furthermore, considering many other
spin-orbital-lattice interacting compounds, such as manganites 
\cite{Shiina97,Mizokawa99,Gu02}, vanadium oxides \cite{Zou05,Tanaka02}, 
and titanium oxides \cite{Cwik03,Schmitz04,Haverkort05,Streltsov05}, 
in which the CF splitting arising from the lattice distortion 
extensively exists, one may find such a fact that the 
low-symmetric CF splittings play crucial roles in singling out 
many degenerate candidates as the sole orbital ordered ground state. 
Thus a conclusion arrives that 
{\it the highly degenerate ground state in the correlated electronic 
system with pure spin-orbital interactions usually stabilizes through 
distorting to a lower symmetry phase},
which is a natural generalization of the Jahn-Teller effect in strongly 
correlated systems. Detail results will be presented in further study.

   In summary, we have performed a systematic study on the roles of the 
electronic SE interaction, the JT orbital coupling and the 
orthorhombic CF splitting in the orbital ordering and the magnetic 
properties in KCuF$_{3}$. The SE and effective JT orbital coupling 
lead to a orbital liquid state due to the inherent frustration and 
orbital quantum fluctuations. The orthorhombic CF lowers the 
orbital symmetry, and stabilizes the orbital ordering as the observed 
in experiment. 
The orbital ordering results in the strong magnetic anisotropy. Strong
spin fluctuation and the orbital frustration considerably reduce
magnetic moment of Cu spins.

\begin{acknowledgments}
\label{acknowledgments}
Authors thanks N. Binggeli for providing the electronic structure data 
of KCuF$_{3}$. Great appreciate is devoted to G. Sandro, M. Fabrizio and 
M. Altarelli in developing the Cluster-SCF method.
Supports from the NSF of China and the BaiRen project from the Chinese 
Academy of Sciences (CAS) are appreciated. Part of numerical calculation 
was performed in CCS, HFCAS.

\end{acknowledgments}

\end{document}